# Channel spaser

A.A. Lisyansky,[1] I.A. Nechepurenko,[2] A.V. Dorofeenko,[2] A.P. Vinogradov,[2] and A.A. Pukhov[2]

[1]Department of Physics, Queens College of the City University of New York, Flushing, NY 11367

[2]Institute for Theoretical and Applied Electromagnetics RAS, 13 Izhorskaya, Moscow 125412, Russia

We show that net amplification of surface plasmons is achieved in channel in a metal plate due to nonradiative excitation by quantum dots. This makes possible lossless plasmon transmission lines in the channel as well as the amplification and generation of coherent surface plasmons. As an example, a ring channel spaser is considered.

Recently, a new branch of quantum optics – quantum nanoplasmonics – has arisen [1, 2]. Advantageous plasmon properties such as small wavelength and a high energy concentration open new perspectives for constructing nano-devices such as waveguides, cavities, and antennas. In this regard, studies of plasmons propagating along 1D objects such as wires [3], wedges [4, 5], and channels [6, 7] are of great interest.

Losses in metals are the main obstacle to practical applications of plasmonics. It has been suggested that this problem can be overcome by compensating loss in a gain medium [8-11]. This relates nanoplasmonics to quantum optics [1, 2]. In particular, a generator of plasmons propagating along a flat surface has been suggested in Refs. [12-14]. In this system, periodic surface corrugations produce Bragg mirrors of the resonator cavity. The first quantum nanoplasmonic device which was referred to as spaser (Surface Plasmon Amplification by Stimulated Emission of Radiation) was proposed in Ref. [1]. The spaser consists of a quantum dot (QD) located near a metal nanoparticle (NP). The plasmonic oscillations in the NP play the role of photons in a laser and the cavity effect is provided by the plasmon localization in the vicinity of the NP [1, 15-17]. Thus, a pumped QD nonradiatively transfers its excitation to surface plasmons localized at the NP. As a result, one observes an increase of intensity of the surface plasmon field. Thus, the spaser does not radiate an energy beam but generates a near field. The spaser has recently been realized [18-21].

In this paper, we study surface plasmons propagating along the bottom of a groove (channel) in the metal surface [22-24] (see schematic in Fig. 1). The surface plasmons are coherently excited by a linear chain of QDs at the bottom of the channel. We show that for realistic values of the system parameters, gain can exceed loss and plasmonic lasing in ring or linear channels can occur.

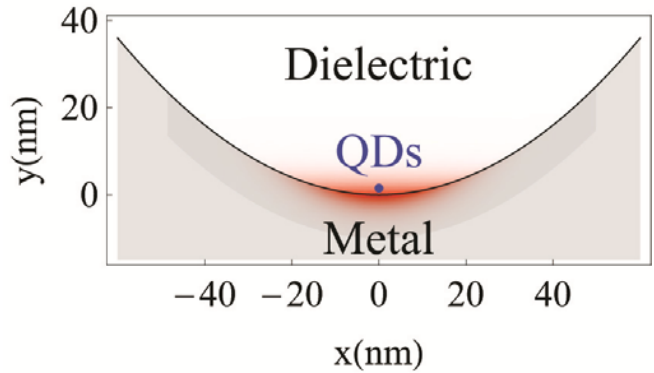

Fig. 1. (Color online) Schematic of the system geometry. The shaded area around QDs shows the electric field intensity distribution calculated for the channel with the bottom curvature of $\rho = 50\,nm$ at $k_z\rho = 10$. The QDs at the bottom of the channel in metal are shown by circles.

In our system, a nanochannel in a metal surface plays the role of a NP in the spaser. As a result, instead of localized plasmons, the travelling plasmons form a beam inside the channel. Utilizing travelling plasmons as a lasing mode differentiates the channel spaser from the spaser proposed by Bergman and Stockman [1]. First, in the channel spaser, radiation is confined in 2D in contrast to the energy localized in a single NP of the Bergman-Stockman spaser. The amplification of radiation propagating along the channel is directional and may be used in designing of a spaser radiating into passive channel sections. Channel-based optical devices are widely discussed in literature [25]. To obtain lasing in a channel spaser one has to organize a cavity inside the channel or employ ring geometry [26]. Second, for the proper choice of the system parameters of the spaser discussed below, the energy is concentrated at the bottom of the channel, and the effectiveness of the interaction of QDs with plasmonic modes is thereby enhanced. Third, in contrast to localized plasmonic oscillations, 1D propagating plasmons exist over a range of frequencies, and therefore, there is no need to tune the QD transition frequency. Unlike known optical schemes, the proposed channel laser works in the frequency range where flat-surface plasmons do not exist, namely, $\varepsilon_M(\omega) + \varepsilon_D > 0$, where $\varepsilon_D$ and $\varepsilon_M$ are the permittivities of the dielectric in the channel and of the metal, respectively [27, 28]. This provides weak electromagnetic coupling between neighboring channels.

Let us consider pumped QDs located at the bottom of a channel (groove) made in a bulk metal sample or in a thick metal film (Fig. 1). In Refs. [6, 29], it was shown that plasmonic modes can propagate along the channel with a wave number $k_z$ larger than that in vacuum, $k_0 = \omega/c$. We are interested in channel plasmons that can be excited by pumped QDs. Besides channel plasmons, QDs may excite plasmons travelling along the flat metal surface. In addition, the QDs may radiate photons. The latter two processes increase losses in the system. However, the excitation of plasmons is more effective than the radiation of photons by a factor $(k_0\rho)^{-3}$, where $\rho$ is a curvature radius of the bottom [30]. Furthermore, the groove geometry suppresses photon radiation, since waves decays exponentially in the groove of a subwavelength width. We therefore neglect the radiation of photons. This approach was justified in Refs. [31, 32] where the radiation of a classical dipole placed near the metallic wire, wedge, and channel was considered. Thus, we assume that the main contribution to radiation losses is from the excitation of 2D surface plasmons which are not localized in the channel. These losses are largely reduced because we choose the working frequency for the channel spaser determined by the condition $\varepsilon_M(\omega) + \varepsilon_D > 0$ for which plasmons at a flat surface do not exist.

Thus, the radiation into the channel plasmons is almost isolated from the external environment, which can be useful for applications in systems, generating and transmitting optical signals along subwavelength channels. However, pumping the QDs can be a problem. We suggest pumping by UV radiation at frequencies exceeding the bulk plasmon frequency in metal.

The metal is then transparent and the UV radiation is effectively absorbed by the QDs and re-radiated at their transition frequency in the luminescence process [33-35].

In view of the above, we neglect the radiative losses and only take into account ohmic losses for plasmons propagation along the bottom of the channel. The amplification of plasmons may be described by the Maxwell-Bloch equations [36-38]:

$$-[\nabla \times [\nabla \times \mathbf{E}]] - \frac{\varepsilon(x,y)}{c^2}\frac{\partial^2 \mathbf{E}}{\partial t^2} = \frac{4\pi}{c^2}\frac{\partial^2 \mathbf{P}}{\partial t^2},$$
$$\frac{\partial^2 \mathbf{P}}{\partial t^2} + \frac{2}{\tau_p}\frac{\partial \mathbf{P}}{\partial t} + \Omega^2 \mathbf{P} = -\frac{2\Omega\mu^2 n\mathbf{E}}{\hbar},$$
$$\frac{\partial n}{\partial t} + \frac{1}{\tau_n}(n - n_0) = \frac{1}{\hbar\Omega}\text{Re}\left(\mathbf{E}^* \cdot \frac{\partial \mathbf{P}}{\partial t}\right),$$
(1)

where $\mathbf{E}$ is the electric field of the channel plasmon, $\varepsilon(x,y)$ is the permittivity of the metal/dielectric system, $\mathbf{P}$ is the polarization of QDs, $n$ is the difference between the concentrations of QDs in the excited and ground states (population inversion), $n_0$ is the equilibrium value of $n$ created by the pumping, $\Omega$ is the transition frequency of the QDs, $\tau_p$ is the inverse line width, $\tau_n$ is the relaxation time of the population inversion, and $\mu$ is the off-diagonal element of the dipole moment of a single QD.

The lasing threshold can be determined in the linear approximation, when the field intensity is so low that population inversion is not suppressed. In other words, one can exclude the 3$^{\text{rd}}$ equation in system (1), and set $n = n_0$. Further, if the plasmon frequency equals the QD transition frequency, $\mathbf{E}, \mathbf{P} \sim \exp(-i\Omega t)$, the polarization can be expressed through the electric field:

$$\mathbf{P} = -i\frac{\tau_p |\mu|^2 n_0}{\hbar}\mathbf{E}.$$
(2)

The gain coefficient of plasmons can be found from the expression for the Poynting vector:

$$\text{div}\,\mathbf{S} = \Omega\,\text{Im}(\mathbf{E}\cdot\mathbf{P}^*) - \Omega\varepsilon''\mathbf{E}\cdot\mathbf{E}^*/4\pi, \tag{3}$$

where $\varepsilon''$ is the imaginary part of permittivity, which reflects losses. We assume that the distances between QDs located at the channel's bottom are small, so that their effect on a plasmon is the same as that of a continuous (constant) gain distribution along the groove ($z$-axis), $n_0 = N_0\delta(x)\delta(y)$, where $x$ and $y$ axes are parallel and perpendicular to the metal surface. Integrating both sides of Eq. (3) over $x$ and $y$ at infinite limits and taking into account Eq. (2), one obtains an expression for the energy flux, $\Phi = \int S_z dxdy$. Setting $\Phi \sim \exp(\gamma z)$ and $|\mathbf{E}(0,0,z)|^2 \sim \exp(\gamma z)$, one obtains the propagation constant:

$$\gamma = \Omega\frac{\tau_p|\mu|^2 N_0}{\hbar}\frac{|\mathbf{E}(0,0,z)|^2}{\Phi(z)} - \frac{\Omega}{4\pi}\frac{\int \varepsilon''|\mathbf{E}(x,y,z)|^2 dxdy}{\Phi(z)}. \tag{4}$$

The rhs of Eq. (4) describes competition between gain and loss in the metal. In the case of plasmons propagating along a flat surface, losses in metal can be compensated by gain [8]. Below, we show that in the case of the channel plasmons, amplification by QDs can exceed absorption in the metal.

Amplification of a plasmon means $\gamma > 0$, which gives the condition on the QD density:

$$N_0 > \frac{\hbar}{4\pi\tau_p|\mu|^2}\frac{\varepsilon''\int_{\text{Metal}}|\mathbf{E}(x,y,z)|^2 dxdy}{|\mathbf{E}(0,0,z)|^2}. \tag{5}$$

Let us evaluate $N_0$ without taking into account the plasmon field redistribution caused by losses and using typical values for QD and channel parameters. In silver, the characteristic value of $\varepsilon''$ is $\sim 0.6$ [39]. For the nanocrystal CdSe QDs $\tau_p \sim 300\,\text{fs}$ and $|\mu| = 20$ Debye [33, 34]. Taking the curvature radius of the channel's bottom to be $\rho = 50\,\text{nm}$ and using an analytical expression for the plasmon field distribution in a channel with parabolic profile [5], we obtain the "effective area" $\int_{\text{Metal}}|\mathbf{E}|^2 dxdy/|\mathbf{E}(0,0,z)|^2 = 75\,nm^2$. One then requires more than $N_0 \sim 30$ excited QDs per micrometer of the channel length to achieve the plasmon amplification. For CdSe QDs of the typical size of a couple of nanometers such a concentration may be easily achieved, even if only one row of QDs is placed at the bottom of channel.

To find optimum conditions for plasmonic lasing, one should analyze the modal structure of radiation. In the limit of large $k_z$, the channel plasmons can be described in the quasi-static approximation [5, 29]. Then the problem is equivalent to that of the quantum oscillator, which localized solutions form a set of discrete levels [5]. Thus, the plasmons form a discrete spectrum (labeled with $\nu = 0,1,2,...$) and are localized at the channel's bottom. Going beyond the quasi-static approximation does not change these conclusions: the channel plasmons are similar to the modes of an ordinary waveguide and form a discrete spectrum. The dispersion curves $\omega(k_z)$ of the lowest-order plasmons corresponding to the lossless case are shown in Fig. 2 (see Refs. [6, 40] for details). The curves in Fig. 2 are calculated within the full-wave approach instead of the quasi-static approximation, because the latter fails at low values of $k_z$ (see Refs. [40] and compare the dash-dotted curve with the solid curve for $\nu = 0$ in Fig. 2).

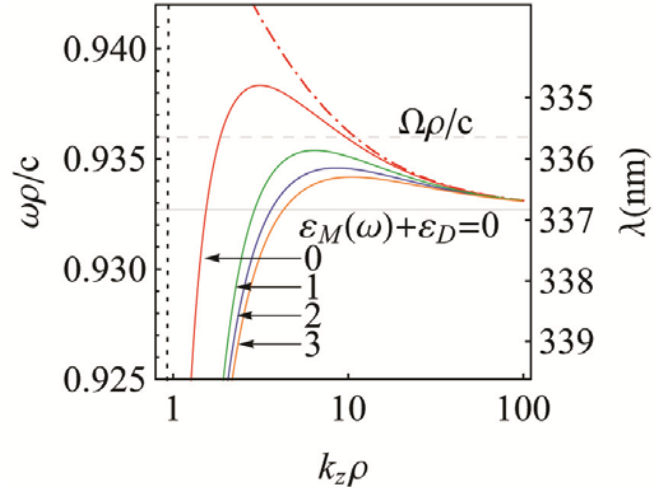

Fig. 2. (Color online) Dispersion curves of channel plasmons, calculated with the account of the retardation effects. The curves are numbered by the plasmon order. The dash-dotted line is the dispersion curve of the lowest-order plasmon, calculated in the quasi-static approximation. The solid horizontal line marks the frequency, at which $\varepsilon_M(\omega) + \varepsilon_D = 0$. The dashed horizontal line corresponds to the transition frequency providing excitation of minimum number of modes. The dots show the light line. The dispersion of silver is taken from Ref. [39], $\operatorname{Im}\varepsilon$ is discarded. The curvature radius of the channel's bottom is $\rho = 50\,\text{nm}$.

The channel plasmon dispersion curves tend to a horizontal asymptote $\omega = \omega_0$ with $\omega_0$ determined by the condition $\varepsilon_M(\omega_0) + \varepsilon_D = 0$. For any given frequency, there may exist plasmonic modes with different wave numbers $k_z$, which form a discrete spatial spectrum. As mentioned above, we consider the frequency range, determined by the condition $\varepsilon_M(\omega) + \varepsilon_D > 0$. As can be seen in Fig. 2, we can choose a frequency $\Omega$ which satisfies both the condition $\varepsilon_M(\omega) + \varepsilon_D > 0$ and the condition of excitation of only a single channel plasmon mode with $\nu = 0$. The lowest-order mode is excited with two different wave numbers $k_z$. The lower and higher values of $k_z$ correspond to forward and backward waves, respectively. The backward wave has a negative slope of the dispersion curve, $\partial\omega/\partial k_z < 0$. This wave while transmitting the energy from the source has a phase velocity towards the source. The backward wave is properly described in the quasi-static approximation, which is confirmed by proximity of the dash-dotted and solid lines in the higher-$k_z$ region in Fig. 2. However, this approximation does not reveal the presence of the forward wave. The plasmon with larger wave number $k_z$ is more localized in the vicinity of the channel's bottom and, thus, has a larger confinement factor $|\mathbf{E}(0,0,z)|^2/\Phi(z)$. As a consequence, this plasmon has a larger gain coefficient in Eq. (4) and a lower threshold value of the QD density $N_0$. Therefore, the channel plasmon with smaller wave number will be suppressed due to mode competition. That is why we consider the higher-$k_z$ channel plasmon in the evaluation of $N_0$. For this plasmon the quasi-static approximation works well and, therefore, the approximation is employed in the calculation of fields.

Since the plasmon amplitude increases while it propagates, lasing appears in a ring channel. Alternatively, one can form an effective cavity in a linear channel. This requires a larger gain level. Thus, ring or linear channels can be employed in a new kind of spaser generating 1D plasmons. The radiation of flat surface plasmons can be suppressed by using the frequency range in which $\varepsilon_M(\omega) + \varepsilon_D > 0$. For the vacuum channel in silver, this range is in the near UV. Replacing the vacuum by a high-permittivity dielectric and silver by a metal-dielectric composite, one can shift the working frequency of the suggested spaser into the visible range.

Indeed, as has been shown in Ref. [41], using either a dielectric matrix filled with silver inclusions or the silver foam with dielectric bubbles, the zero-epsilon frequency $\omega_{zero}$ ($\mathrm{Re}\,\varepsilon_{eff}(\omega_{zero}) = 0$) may be significantly shifted below the plasma frequency of silver.

**Acknowledgements**

The authors are grateful to A.V. Fedorov and A.V. Baranov for discussion of the properties of quantum dots. The work is partly supported by the Dynasty Foundation, by the RFBR projects 10-02-00857-a, 10-02-90466-Ukr_a, 10-02-91750-AF_a, 11-02-92475-MNTI_a and by PSC-CUNY grant.